\documentstyle[aps,preprint,tighten,floats,multicol,epsf,rotate]{revtex}

\begin{document}
\draft
\title{Stochastic Resonance in Underdamped, Bistable Systems}
\author{Rajarshi Ray$^{a,b}$ \footnote{E-mail: rajarshi.ray@saha.ac.in} and 
Supratim Sengupta$^{c,d}$ \footnote{E-mail: sengupta@physics.mcmaster.ca}}
\address{\it $^a$Tata Institute of Fundamental Research, Homi Bhabha Road, 
Mumbai 400 005, India.\\ 
$^b$Thoery Group, Saha Institute of Nuclear Physics, 1/AF, Bidhan Nagar
Kolkata 700 064 India.\\
$^c$ Department of Physics and Astronomy, Mcmaster University, Hamilton 
L8S 4M1, Canada.\\
$^d$Theoretical Physics Institute, Department of Physics, University of 
Alberta, Edmonton T6G 2J1, Canada.}
\date{\today}
\maketitle
\widetext
\parshape=1 0.75in 5.5in
\begin{abstract}

 We carry out a detailed numerical investigation of stochastic resonance in 
underdamped systems in the non-perturbative regime. We point out that an 
important distinction between stochastic resonance in overdamped and 
underdamped systems lies in the lack of dependence of the amplitude of the 
noise-averaged trajectory on the noise strength, in the latter case. We 
provide qualitative explanations for the observed behavior and show that 
signatures such as the initial decay and long-time oscillatory behaviour 
of the temporal correlation function and peaks in the noise and phase averaged power 
spectral density, clearly indicate the manifestation of resonant behaviour in noisy, 
underdamped bistable systems in the weak to moderate noise regime. 
\end{abstract}
\vskip 0.125 in
\parshape=1 -.75in 5.5in
\begin{multicols}{2}
\narrowtext
 
\section{Introduction}
 
  Noise usually has a disruptive role in nature. However, there is a
phenomenon \cite{benzi,nicolis} in which noise helps in making the system behave 
in a more coherent manner. This intriguing and rather counterintuitive 
phenomenon called {\it Stochastic Resonance}(SR) \cite{rmp} was 
first proposed as an explanantion of the observed periodicity in the 
ice ages on earth \cite{benzi,nicolis}, but has since been observed in a large
variety of physical, chemical and biological systems \cite{moss}. Bistable systems 
subject to noise and periodic modulation constitute the most common class of systems
which exhibit SR. 

 The phenomenon of SR can be thought of as resulting from the synchronization
of a stochastic time-scale (determined by the thermal transition rate over
the barrier) and a deteministic time scale (determined by the time-scale of
periodic modulation). When the noise strength is appropriately tuned to ensure
the synchronization of the these two time scales, the system exhibits stochastic
resonance. SR manifests itself through a variety of signatures such as peak in
the amplitude of the noise-averaged trajectory versus noise strength, peak in the 
distribution of peak heights in the 
residence time distribution (RTD) \cite{march1,zhou,fox}, peaks in the power spectral 
density (or Signal-to-Noise Ratio) \cite{mcnamara} at the modulating frequency and 
late-time oscillatory behaviour of the temporal auto-correlation function \cite{march1}. 
SR has been extensively investigated in many overdamped systems, but it has'nt been 
explored in great detail in underdamped systems. (references listed in \cite{march1,alfonsi} 
are some exceptions). Apart from the intrinsic interest in exploring in detail 
the manifestations of SR in the under-damped regime, there are other motivations 
for focussing on this regime. In particular, relativistic systems are governed 
by underdamped equations of motion and investigating SR in single particle underdamped 
systems may provide valuable insights into the manifestations of SR in relativistic 
field theories. 

 We will focus on numerically investigating the phenomenon of 
SR in underdamped bistable systems subject to white noise and periodic 
modulation. The system we are dealing with can be described by the Langevin equation 
\begin{eqnarray}
&&\frac{d^2x}{dt^2} + \eta\frac{dx}{dt} - x + x^3 = A_ocos(\omega t + \theta)
+ \xi(t) \nonumber  \\
&&\langle \xi(t)\xi(t^{\prime}) \rangle = 2T\eta\delta(t-t^{\prime}) \label{eq1}
\end{eqnarray}
where $T$ is the temperature of the background heat bath, $\eta$ is the dissipation
coefficient, $A_0$ and $\omega$ are the amplitude and frequency of periodic modulation 
repectively. $\xi$ is a Gaussian distributed (delta correlated) white noise
term which satisfies the fluctuation-dissipation theorem. For simplicity, we have 
made use of scaled, dimensionless quatities in the above equation. The potential 
governing particle dynamics is of standard double well type, modulated by a
periodic signal with small amplitude. 
\begin{eqnarray}
V(x,t) = -\frac{1}{2}x^2 + \frac{1}{4}x^4 - A_0x\cos(\omega t + \theta)
\end{eqnarray}
In the absence of periodic modulation, the doubly degenerate minima of the 
double well potential are located at $x_m=\pm1$ and the barrier height is 
given by $\Delta V_0 = 0.25$ The modulation amplitude $A_0$ is small enough to 
ensure the bistability of the potential at all times. Moreover, the noise strength, 
$D=T\eta$ (for the underdamped case) is much smaller than the barrier height in
the absence of modulation i.e. $D << \Delta V_0$. We will also be working in
the moderately weak noise regime $D\lesssim A_0$. There has been several analytical 
studies of SR in overdamped systems \cite{rmp}, but those investigations were 
based on perturbation theory and therefore restricted to the perturbative regime 
$A_0<<D$ or the regime of validity of the adibatic approximation $A_0 << D << \Delta V_0$. 
An analytical study of underdamped, periodically modulated systems was carried out
by Jung and Hanggi \cite{hanggi}, based on the solution of the 2-dimensional and
3-dimensional Fokker-Planck equations for probability distribution. They obtained 
the asymptotic probability distribution and showed that for uniformly distributed
initial phases, the power spectrum exhibits delta-functions spikes at 
multiples of the driving frequency. Previous experimental work on SR in underdamped 
systems \cite{march1} have been restricted to the strong noise regime characterized 
by $T\eta >> \Delta V$ and focussed primarily on signatures like the 
signal-to-noise (SNR) ratio and residence time distribution. More recently,
Alfonsi et.al. \cite{alfonsi} have shown that that an intra-well SR can coexsist
with conventional inter-well SR in underdamped systems and is manifest as 
twin peaks in the plot of power spectral density versus noise strength. 
Their paper further extended the work of Stocks et.al. dealing with SR in monostable 
systems \cite{stocks} using methods of Linear Response Theory. Evidence of SR
in {\it overdamped} bistable systems without explicit symmetry breaking and subject
to periodically oscillating noise strength have been discussed in \cite{neiman,gammai}.

 Our main aim in this paper is to numerically explore the various manifestations of 
SR in {\it underdamped} systems in the moderately weak noise and moderate forcing regime. 
This is distinct from the strong noise, weak forcing regime \cite{march1} and strong 
forcing regime \cite{alfonsi} explored in earlier works. Although the effect of weak 
noise was considered in the work of Alfonsi {\it et.al.}\cite{alfonsi}, their analysis 
was focussed on the behaviour of the power spectrum and in distinguishing between intrawell 
and interwell SR in the high frequency forcing regime. 
We point out an unusual behaviour of the noise-averaged trajectory, in 
under-damped systems in the moderate-weak noise and weak forcing regime, 
which is quite distinct from its behaviour in over-damped systems
where it is widely used as a characteristic signature of SR. We find that there is no 
increase in the amplitude of noise averaged trajectory with increasing noise strength, 
unlike SR in overdamped systems. This important distinction in the dynamics of overdamped 
and underdamped systems was not recognized earlier. It can be attributed to the delicate 
balancing act between noise and dissipation in underdamped systems, which also reduces the 
window size in parameter space for observing SR-like behaviour. Our results supplement
the analysis of SR in underdamped systems carried out earlier \cite{march1,alfonsi}.

    The paper is organized as follows. In Sec. II, we carry out a detailed numerical study 
of SR in underdamped systems in the weak noise, moderate forcing regime. The effect of 
changing system parameters on the various distinct signatures of SR are described in detail. 
We conclude with a brief summary and discussion of our results in section III. 

\section{Stochastic Resonance in Underdamped Systems : Moderate and Weak Noise Regime}

 The key idea behind SR is that the synchronization of the stochastic and deterministic
time scales of the system can result in a more coherent response of the system even on 
{\it increasing} the noise strength, for a small range of values of the modulating 
frequency. Even though SR was first investigated for overdamped systems, this intuitive
notion of SR was also found to be valid for underdamped systems \cite{march1}. The 
interesting area of parameter space is such that neither the periodic modulation of the 
potential, nor the thermal noise activation is individually capable of ensuring 
periodicity in the inter-well hopping of the particle. 
For large noise strengths i.e. $D >> \Delta V_0$ 
the particle has a large enough thermal energy to easily surmount the potential 
barrier and therefore does not feel the presence of the barrier. In the 
absence of coupling to a heat bath, the small amplitude of modulation, 
$A_0 < \Delta V_0$ and forcing frequency $\omega << \tau_{r}^{-1}$ 
(where $\tau_r$ is the intra-well relaxation time scale) cannot provide 
sufficient energy to the particle to surmount the barrier. However, both these effects
acting in conjunction can, under certain circumstances, lead to periodic behaviour
in the noise-averaged particle trajectory. In general, the thermal transition rate 
(Kramer's rate) over the barrier $\tau_k^{-1} \equiv r_k $, depends on both the 
noise strength and dissipation coefficient. 
The time scale of periodic modulation ($\tau_0=\frac{2\pi}{\omega}$) is determined 
by the forcing frequency $\omega$. Due to the odd power of the coupling between the 
particle position and the periodic forcing, the symmetry of the potential 
is explicitly broken. The particle in either well encounters the smallest 
(largest) barrier twice during a single period. If $\tau_k^{(-)}$  
($\tau_k^{(+)}$) are the corresponding thermal transition time-scales, 
the noise averaged particle trajectory becomes periodic when  
$\tau_k^{(-)} < \tau_0 < \tau_k^{(+)}$.
Inter-well hopping is therefore synchronized when the following (approximate)
condition between the deterministic and stochastic time scales is satisfied 
$\tau_k \simeq \frac{\tau_0}{2}$ $\Rightarrow$ $\omega_{sr} \sim \frac{\pi}{\tau_k}$.
Fox and Lu \cite{lu} have shown that this matching condition which yields the 
corresponding resonant value of the noise strength ($D_{sr}=(T\eta)_{sr}$), 
is an approximate one and does not become exact in any limit. 

 At this stage it is necessary to point out the crucial difference between our
results and those obtained via analog simulations \cite{march1} in the work 
of Gammaitoni et.al. The main difference lies in the 
distinctly different regime of study undertaken in this paper. Even though
the work of Gammaitoni et.al. \cite{march1} dealt with investigating SR in 
underdamped systems, their choice of parameters restricted their
study to the strong noise regime characterized by the condition 
$T\eta > \Delta V_0$. However, we decided to explore the weak-to-moderate
noise regime characterized by the condition $T\eta << \Delta V_0$ and 
$T\eta \lesssim A_0$. Moreover, the validity of their theoretical 
analysis depended on the applicabiliy of perturbation theory, which amounts 
to the  condition $A_0 << T\eta$ implying that the influence of modulation
is small compared to thermal activation. This lead to the temperature dependence 
of the amplitude of the noise-averaged trajectory. However, our regime of study is
not amenable to a perturbative analysis since in our case $T\eta \lesssim A_0$.
Moreover, $A_0 \lesssim \Delta V_0$ further implying a substantial (non-perturbative) 
modulation of potential by the driving force, even though the modulation is 
insufficient to destroy the bistability of the potential at any time and is therefore 
distinct from regime where resonant trapping \cite{apostolico} is observed. We find that 
this leads to an important difference in the behaviour of the noise averaged trajectory 
characterized by the lack of dependence of the amplitude with noise strength. This is 
distinct from the manifestation of SR in the overdamped case and in the underdamped 
case studied earlier \cite{march1}.

 The particle dynamics in the moderate-strong damping regime is dominated by fluctuations and 
the Kramer's rate in this regime is given by
\begin{equation}
r_k = \frac{\sqrt{1+\eta^2/4}-\eta/2}{\sqrt{2}\pi}e^{-\beta\Delta V_0}
\end{equation}

However, this expression is derived under the assumption that the intrawell relaxation 
time-scale of a particle is much smaller than the thermal transition time scale and the 
particles are able to thermalize before escaping over the barrier. As the dissipation 
coefficient decreases, this assumption breaks down as the motion of the particles starts 
being dominated by energy diffusion. Appropriate modification to the rate formula 
in the very weak dissipation regime was obtained by Kramers \cite{kramer,talkner}, 
Carmeli {\it et.al.} \cite{carmeli} and by Buttiker {\it et.al.} \cite{buttiker},
in which the transition rate was found to be proportional to $\eta$

\begin{equation}
r_k = \frac{\eta\beta I(E_0)}{2\sqrt{2}\pi}e^{-\beta\Delta V_0}
\end{equation}

\noindent where $I(E_0)$ is the action at the top of the barrier. [Carmeli {\it et.al.} \cite{carmeli} 
and Buttiker {\it et.al.} \cite{buttiker} also obtained appropriate corrections to the 
above formula by relaxing the approximation that the distribution function drops to zero at 
the top of the barrier, a crucial assumption in the derivation of Eq.(4).] The approximate 
resonant frequency $\omega_{sr} \sim \frac{\pi}{\tau_k}$ for underdamped systems has to be 
estimated with this modified rate formula in the very weak dissipation regime. In our work, 
$\omega_{sr}$ was used only to identify the regime for resonant behaviour in underdamped 
systems.

   We numerically solved Eq.(\ref{eq1})using a second order stochastic Runge-Kutta 
algorithm.  The noise averaged trajectory was obtained by averaging over one hundred
noise realizations. The upper plot of Fig.1 shows the noise averaged trajectory for 
two different values of bath temperature, obtained by solving the overdamped Langevin 
equation given by 
\begin{eqnarray}
&&\frac{dx}{dt} - x + x^3 = A_ocos(\omega t)
+ \xi(t) \nonumber \label{eq2} \\
&&\langle \xi(t)\xi(t^{\prime}) \rangle = 2T\delta(t-t^{\prime})
\end{eqnarray}
A clear dependence on the amplitude of the oscillation on temperature is observed.
The maximum of the amplitude corresponds to the temperature that satisfies the  
resonant condition $\omega_{sr} \sim \frac{\pi}{\tau_k} $.
When the temperature increases beyond 0.08, the amplitude decreases and eventually the 
dynamics becomes noise dominated resulting in loss of periodicity. In contrast, the lower 
plot of Fig.1 shows the noise averaged trajectories for underdamped case obtained 
by solving Eq.(\ref{eq1}). The trajectories for the two different temperature values 
almost overlap indicating a lack of dependence on the temperature of the heat bath. 
This important aspect of SR in underdamped system in the weak-moderate 
noise regime was not realized earlier. It can be attributed to the fact that 
the noise strength ($T\eta$) in underdamped systems depend not only on the 
temperature of the heat bath, but also on the dissipation coefficient $\eta$. 
Since $\eta$ is very small for underdamped systems, an increase in temperature 
is offset by the smallness of $\eta$ which therefore does not lead to any 
substantial increase in noise strength. 

\subsection{Residence Time Distribution}

 An useful characterization of SR \cite{march1} is obtained by 
considering the discrete stochastic process associated with the barrier crossing time.  
The residence time then corresponds to the time spent by the particle in a well 
between successive transitions over the barrier and is obtained by taking the 
difference between successive barrier crossing times. The residence time was obtained 
numerically by noting the time required by the particle starting from either well to cross 
a point specified on the other side of the barrier for each trajectory. The average of 
these times over all realizations yielded the mean residence time. The choice of the 
crossing point, although arbitrary, was made to ensure that the particle had indeed made a 
transition from one well to another. This rules out trajectories where the particle just 
crosses the barrier, lingers around 
the top of the barrier, before falling back into the well it started from. This phenomenon is 
particularly relevant in underdamped systems and would lead to underestimating the 
residence time. We emphasize that the two-state approximation was not been made in estimating 
the residence time. Although it is often customary to make such an approxiation in analyzing 
hopping dynamics in overdamped systems, it would lead to wrong estimates of the residence time 
for reasons mentioned above. Fig.2 shows the residence
time distribution for 3 different values of the bath temperature. A pattern of 
peaks in the RTD with exponentially decaying peak heights is observed. (The peak 
height at $t/\tau=0.5$ have been normalized to unity in Figs.2,4,6.) The likelihood 
of the particle hopping over the barrier is largest when the barrier height
to be surmounted is the least. Due to the modulation of the potential, the 
particle encounters the smallest barrier twice during every modulation 
period. When perfect synchronization is achieved i.e. the resonant condition is 
satisfied, the particle hops over the barrier {\it twice} every modulation period 
and a single large peak in the RTD is observed at $t=0.5\tau_0$. When $T>T_{sr}$
(where $T_{sr}$ is the bath temperature which corresponds to the resonant frequency
$\omega_{sr}$ for fixed $\eta$), noisy dynamics starts dominating and synchronization 
of inter-well hopping is lost. The inter-well transitions occur for $t<0.5\tau_0$, 
which accounts for a sharp peak in RTD for $t<0.5\tau_0$. For $T<T_{sr}$, perfect 
synchronization is lost because the thermal transition time scale($\tau_k$) is 
larger than $0.5\tau_0$. If the particle in unable to traverse the barrier when the 
barrier height is minimum, it has to wait for one full period before it again faces 
the smallest barrier. This accounts for the multiple peaks in the RTD at {\it odd} 
multiples of $0.5\tau_0$.  

 We would like to emphasize that the role of dissipation on SR can only be 
addressed by studying UD systems. This is easily understood since the thermal 
transition time scale for inter-well hopping depends on the $\eta$, albeit weakly. 
However, the signature of SR associated with the periodicity of the noise-averaged 
trajectory is not very sensitive to changes in $\eta$ and $\omega$. This can be seen 
from the plots of the noise averaged trajectory for different values of $\eta$ and 
$\omega$ as shown in Fig.3 and Fig.5. The periodicity in the noise averaged trajectory 
is observed even when $\eta$ is increased by about two orders of magnitude. A more 
sensitive indicator of resonant behaviour is the residence time distribution. 
Nevertheless, the dependence on $\eta$ is rather weak as is evident from the plots 
for the RTD shown in Fig.4. Increasing $\eta$ increases the noise strength thereby 
increasing fluctuations, but that increase is balanced out by the damping 
of fluctuations resulting from a larger $\eta$. Because of this delicate balance 
between fluctuation and dissipation observed in underdamped systems, the dependence 
of RTD on $\eta$ is weaker than its dependence on the bath themperature. This is evident 
from Fig.4 which shows plots of the RTD for three different values of $\eta$. For very 
small dissipation, $\tau_k$ becomes larger than $0.5\tau_0$ and this leads to loss of 
synchronization of inter-well transitions resulting in multiple peaks in RTD at odd 
multiples of $0.5\tau_0$. However, the weak dependence on $\eta$ does not affect the 
shape of the RTD on increasing $\eta$ by an order of magnitude (compared to the 
benchmark value of $\eta=0.03$) as is evident from the last two plots of Fig.4. 

 Fig.5. shows the the noise averaged trajectory for three different values of the 
modulating frequency. The periodic behaviour is observed even for values of $\omega$
substantially different from the value which satisfies the resonant condition. 
However, as $\omega$ is increased, synchronization is lost and the dynamics becomes
noise dominated as is evident from the last plot of Fig.5. However the sensitivity
to change in $\omega$ is more pronounced in the plots for the RTD shown in Fig.6.
When perfect synchronization is attained, inter-well hopping over the barrier occurs 
when the the barrier height is minimum. This amounts for a single sharp peak in the 
RTD at $t=0.5\tau_0$ corresponding to the resonant frequency $\omega_{sr}=0.03$, as 
seen in the first plot of Fig.6. For larger (or smaller) modulation frequencies, synchronizations 
is lost and this is manifest through appearance of multiple peaks (with exponentially 
decreasing peak heights) in the RTD at odd multiples of the half-period of modulation. 

 However, a proper signature of SR based on RTD is characterized by a peak in the 
distribution of peak heights (at half the driving period) versus noise \cite{zhou}.
We also find a peak around $T_{sr}$, in the distribution of peak heights plotted 
against temperature. However, no such peak is observed when the distribution of 
peak heights are plotted against the dissipation coefficient, providing another 
indication that the hopping dynamics is less sensitive to dissipation coefficeint in the 
weak damping regime. This characterization of SR has lead to some controversy regarding 
the identification of SR as a bonafide resonance \cite{fox,march2,dan,sekimoto} in the 
overdamped case. 
Choi et.al. \cite{fox} have argued that for large $A_0$/D values, peaks in the RTD distribution 
were insufficient to imply resonant behaviour in a practically viable frequency range.
Our aim in carrying out the RTD analysis for varying temperature, dissipation and driving 
frequency was to indicate that the behaviour of the RTD in underdamped systems follows the 
same pattern seen in overdamped systems and can be understood in terms of synchronization 
of hopping over the barrier. However, since we are working in the large $A_0$/D regime, RTD alone 
cannot be used to infer about the manifestation of SR in underdamped systems. 
In order to unambiguously show that SR is observed in underdamped systems in the 
weak and moderately weak noise regime (i.e.$A_0/(T\eta) >> 1$) investigated in this paper, 
we present below two other signatures of SR.

\subsection{Auto-correlation Function and Power Spectrum} 

  Evidence for SR can also be obtained by studying the behaviour of the 
auto-correlation function (ACF) \cite{march1}, defined by the equation
\begin{equation}
C(t) = \langle x(t)x(0) \rangle
\end{equation}
where the $\langle...\rangle$ implies averaging over noise realizations. 
The first plot in Fig.7 shows the early-time exponential decay of 
the ACF for four different values 
of the bath temperature. Larger bath temperatures lead to a faster decay in 
the ACF, $C(t)$, since large fluctuations destroy correlations at large 
temporal separations.  The initial decay of the ACF is followed by 
oscillatory behaviour with a frequency equal to that of the the modulation
frequency. Such a behaviour is symptomatic of SR in driven, diffusive systems 
subject to thermal noise. The lower graph of Fig.7 shows a distribution 
of the temporal ACF at late times which indicates a clear oscillatory dynamics.
The gray solid line is a sinosoidal fit to the numerical ACF data and has 
a frequency equal to the modulation frequency of the system. Note that 
our choice of parameters does not allow for a perturbative analysis which was
carried out in \cite{march1} to obtain anaytical expressions for the amplitude
of the noise averaged trajectory and the auto-correlation function. Such 
an approximation is only valid 
for small perturbations
$A_0x_m << T$. \footnote{In the notation of \cite{march1}, $D$ corresponds to 
the temperature ($T$) of the heat bath.} However, in our case, 
$T \gtrsim \Delta V_0$. Furthermore, our regime of study lies {\it beyond}
the perturbative regime, since $A_0x_m \lesssim T$.    

 Another common signature of SR can be obtained by looking at the 
noise and phase averaged power spectral density  $S(\omega)$ obtained by 
taking the fourier transform of the ACF $C(t)$. 
\begin{equation}
S(\omega) = \int_{-\infty}^{\infty}\exp^{-i\omega t}<<x(t+\tau)x(t)>> d\tau
\end{equation}
where the inner brackets indicate avergaing over noise realizations and the outer
brackets denote avergaing over the initial random phase. We averaged over 50 different 
noise realizations and 20 different values of random initial phase to first 
obtain the noise and phase averaged temporal ACF for 64 different values of 
temporal separation. The fourier transform of the noise and phase averaged ACF then 
yielded the noise and phase averaged power spectral density. Fig.8 shows the power 
spectral density as a function of the frequency, obtained for four different values of 
the bath temperature, all other parameters remaining fixed. The plots clearly show a 
sharp peak at the the modulation frequency, and indicates that the peak strength at the 
driving frequency goes through a maximum as a function of the noise strength. Although,
the exact location of the maxima is difficult to ascertain precisely, it is nevertheless 
clear that the maxima is close to $T=0.2$, which is the temperature at which the synchronization 
condition is satisfied for the given choice of parameters. 
These additional signatures clearly establish that the observed synchronization, in 
the weak and moderately weak noise regime of underdamped bistable systems, is indeed a 
manifestation of stochastic resonance.

\section{Conclusions}
 To summarize, we have carried out a numerical study of SR in underdamped,
bistable systems in the regime characterized by the conditions 
$T\eta << \Delta V_0$ and $T\eta << A_0x_m$. 
This is distinct from the regime investigated earlier \cite{march1} and 
is not amenable to an analysis based on treating the modulation of the potential as
a small perturbation. A comparison between overdamped and underdamped 
dynamics was carried out. No significant dependence of the amplitude of 
the noise averaged-trajectory, on the 
temperature of the heat bath, was found in the underdamped scenario.
This is in marked contrast to the the overdamped case, where the amplitude
depends sensitively on the bath temperature. Synchronization of hopping with 
periodic modulation of the potential well was observed in the behavior of the RTD,
as in the overdamped case. The RTD  was found to be more sensitive to changes in 
bath temperature and only weakly dependent on changes in the dissipation coefficient. 
The auto-correlation function and the power spectral density were also obtained for 
various values of the bath temperature to unambiguously confirm that the underdamped 
system does indeed exhibit resonant behaviour in the weak noise and weak forcing regime. 
The power spectrum was characterized by a sharp peak at a value close to that of the 
modulating frequency, with the largest peak height corresponding to the temperature 
at which the resonant condition was satisfied. These signatures provide evidence
that SR in underdamped systems depends on the interplay between the
bath temperature and the dissipation coefficient and leads to a rather
distinct pattern of dynamics, not observed in the previously 
investigated regimes of parameter space.   

 An interesting extension of this work would involve looking for 
evidence of SR in relativistic field theoretic models. The study of 
stochastic resonance for spatially extended systems has been carried 
out for Ginzburg-Landau type field theories \cite{sutera}, but has been 
restricted to the over-damped regime .There it was found 
that an appropriate choice of the frequency of the periodic driving 
obtained by matching the thermal activation time-scale to half the period 
of the modulating background, can result in periodically synchronized behavior 
of the mean field about $\phi =0$ (see Fig.2 of Ref. \cite{sutera}). It would
be interesting to explore the possibility of observing SR in {\it underdamped}, 
field-theoretic models exhibiting spontaneous symmetry breaking of a 
$Z_2$ symmetry since it would provide the first evidence of emergence of 
coherent behaviour in noisy, underdamped field theory. 

\section{Acknowledgements}

 We would like to thank A.M. Srivastava for useful discussions and
encouragement. SS would also like to thank F.C. Khanna for support and 
encouragement. The work of SS was funded in part by NSERC, Canada. 
 

\end{multicols}

\newpage

\vskip -0.25in
\begin{figure}[h]
\begin{center}
\leavevmode
\epsfysize=18truecm \vbox{\epsfbox{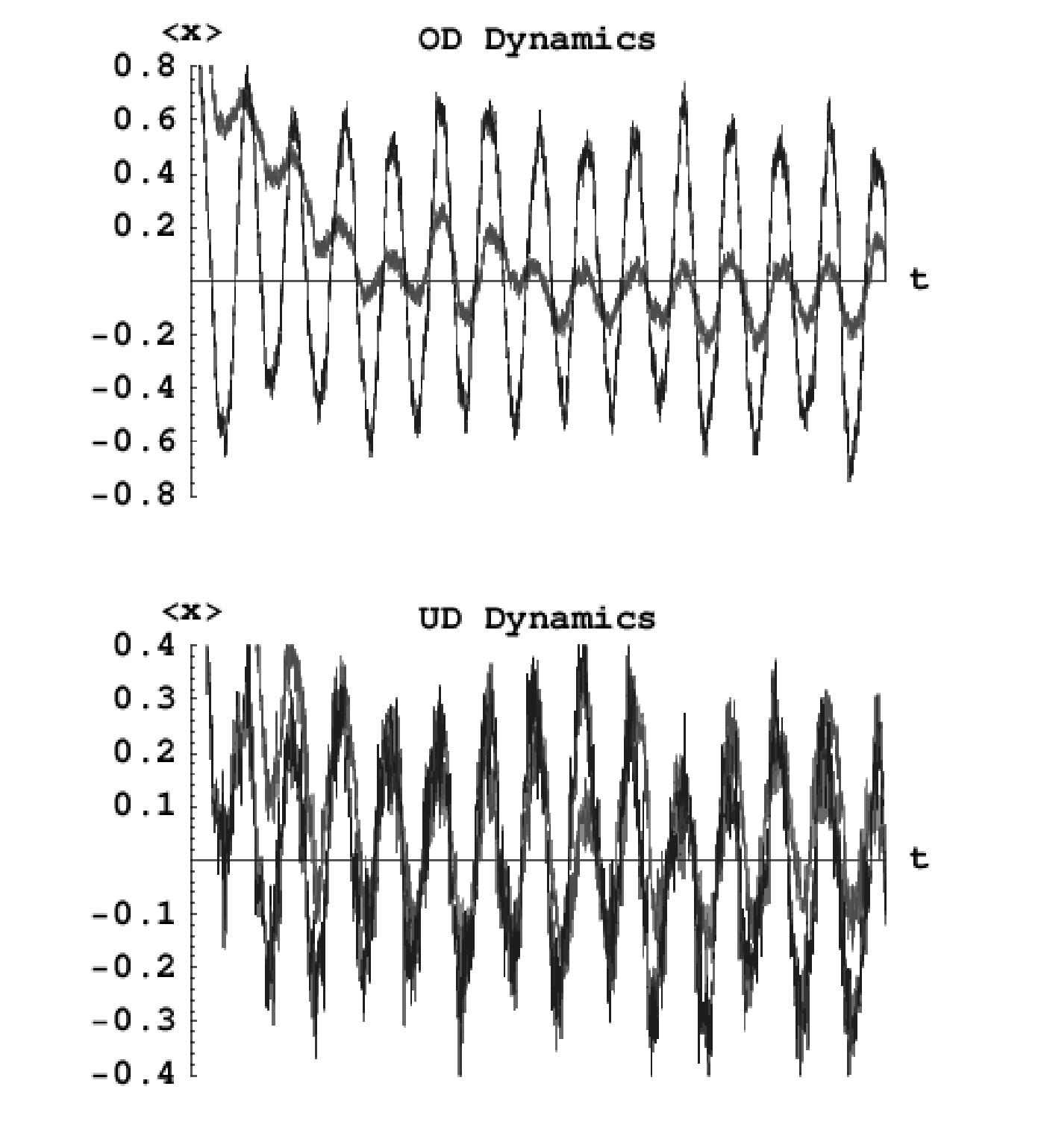}}
\end{center}
\caption{Plots showing the noise averaged trajectory with 
two different temperatures for the overdamped case (top) and underdamped 
case (bottom).The bath temperatures for the two trajectories shown correspond 
to $T=0.04,0.08$ (top) and $T=0.08,0.2$ (bottom).} 
\label{Fig.1}
\end{figure}

\newpage

\vskip -0.45in
\begin{figure}[h]
\begin{center}
\leavevmode
\vskip -0.35in
\epsfysize=20truecm \vbox{\epsfbox{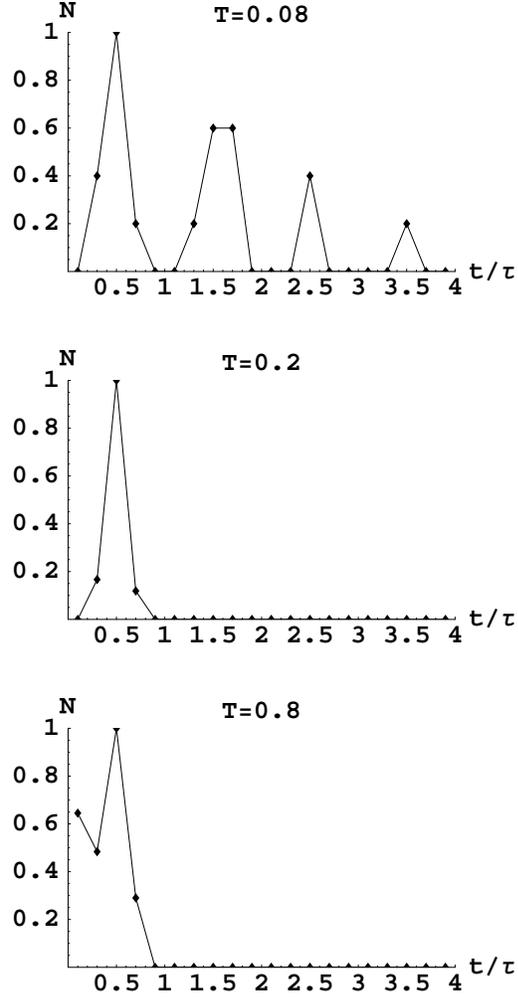}}
\end{center}
\vskip -1.5in
\caption{Residence Time Distribution for three different values of 
bath temperature. Other fixed parameter values are $\eta=0.03, A_{0}=0.1, 
\omega=0.03$. The temperature at which the resonance condition is satisfied
for fixed $\omega$ and $\eta$ is $T \equiv T_{sr}=0.2$}
\label{Fig.2}
\end{figure}

\newpage

\vskip -0.45in
\begin{figure}[h]
\begin{center}
\leavevmode
\vskip -0.35in
\epsfysize=20truecm \vbox{\epsfbox{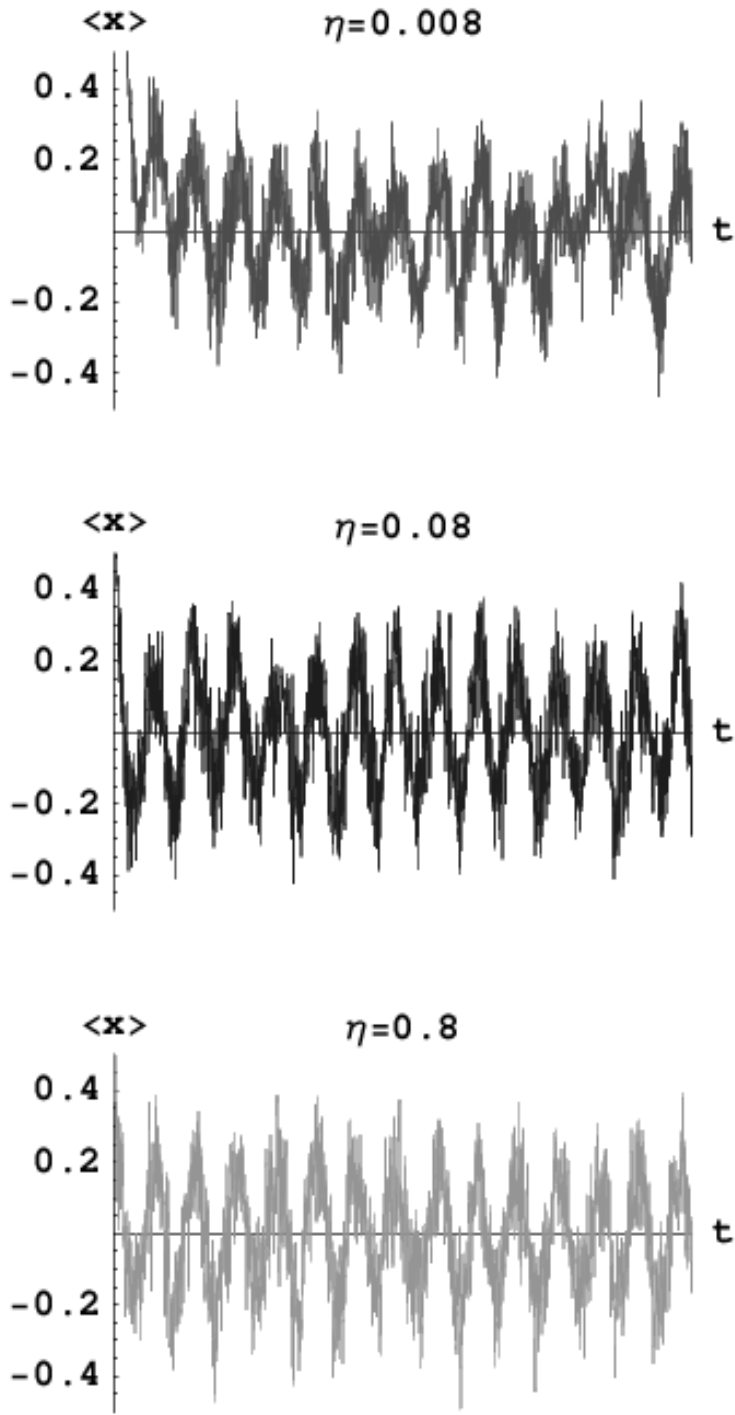}}
\end{center}
\caption{Noise averaged trajectory for three different values of 
$\eta$. Other fixed parameter values are $T=0.2, A_{0}=0.1, 
\omega=0.03$. The dissipation at which the resonance condition is satisfied
for the given $\omega$ and $T$ value is $\eta \equiv \eta_{sr}=0.03$}
\label{Fig.3}
\end{figure}

\newpage

\vskip -0.45in
\begin{figure}[h]
\begin{center}
\leavevmode
\vskip -0.35in
\epsfysize=20truecm \vbox{\epsfbox{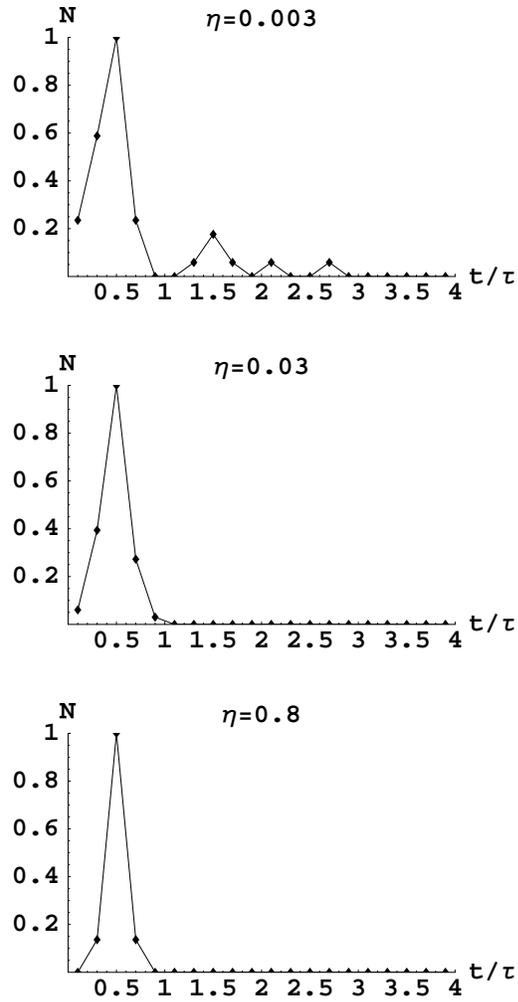}}
\end{center}
\vskip -1.5in
\caption{Residence Time Distribution for three different values of 
$\eta$. Other fixed parameter values are $T=0.2, A_{0}=0.1, \omega=0.03$.}
\label{Fig.4}
\end{figure}

\newpage

\vskip -0.5in
\begin{figure}[h]
\begin{center}
\leavevmode
\vskip -0.5in
\epsfysize=20truecm \vbox{\epsfbox{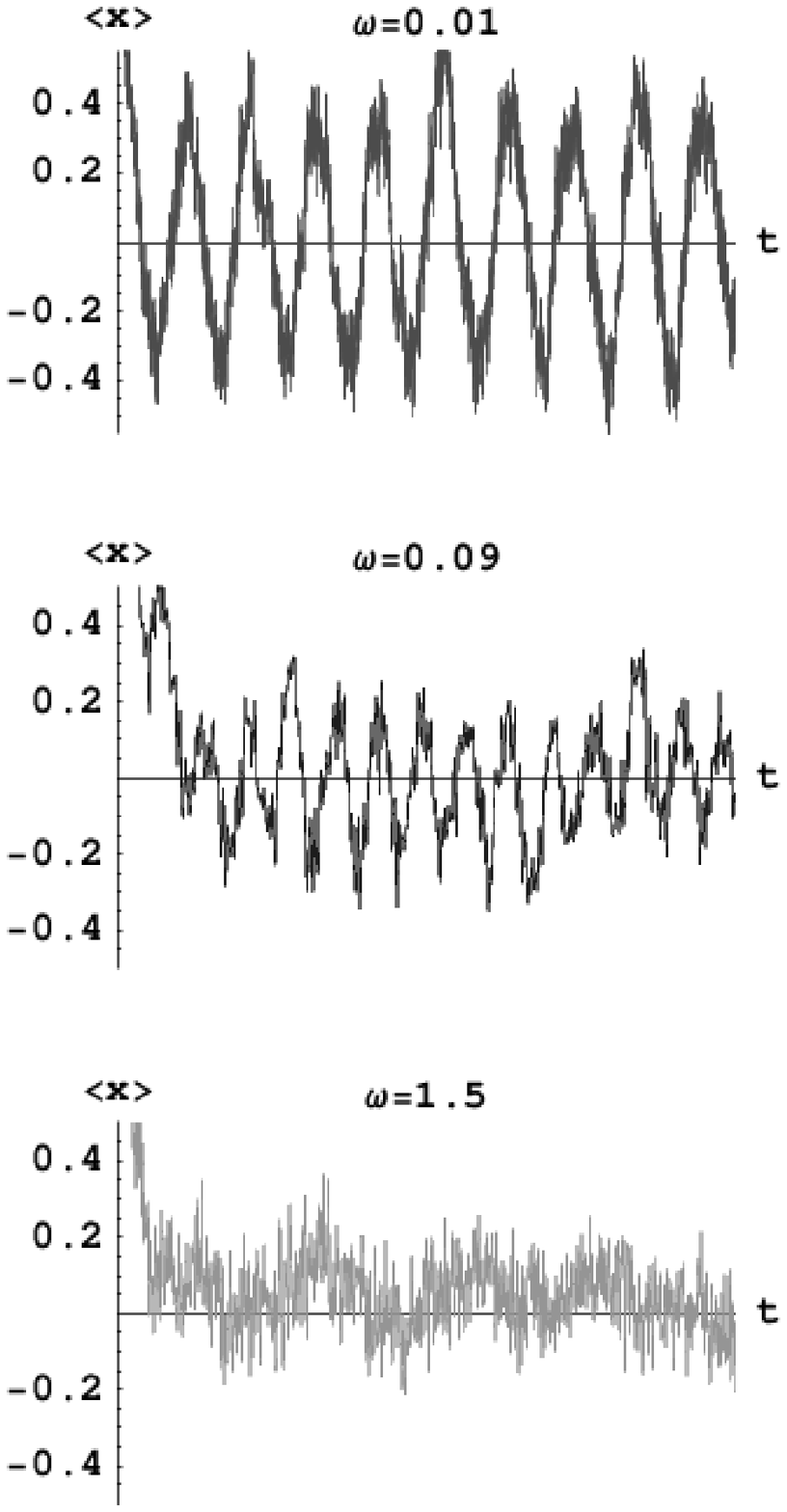}}
\end{center}
\caption{Noise averaged trajectory for three different values of the
modulating frequency. Other fixed parameter values are $T=0.2, A_{0}=0.1, 
\eta=0.03$. The resonant frequency for the given $\eta$ and $T$ values 
correspond to $\omega \equiv \omega_{sr}=0.03$} 
\label{Fig.5}
\end{figure}

\newpage

\vskip -0.5in
\begin{figure}[h]
\begin{center}
\leavevmode
\vskip -0.5in
\epsfysize=20truecm \vbox{\epsfbox{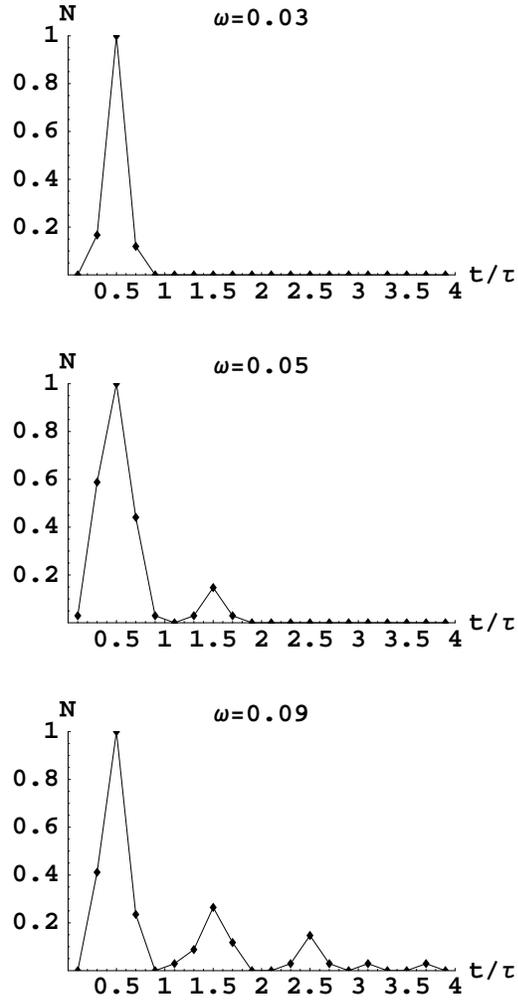}}
\end{center}
\vskip -1.5in
\caption{Residence Time Distribution for three different values of the
modulating frequency. Other fixed parameter values are $T=0.2, A_{0}=0.1, 
\eta=0.03$. The resonant frequency for the given $\eta$ and $T$ values 
correspond to $\omega \equiv \omega_{sr}=0.03$} 
\label{Fig.6}
\end{figure}

\newpage

\vskip -0.35in
\begin{figure}[h]
\begin{center}
\leavevmode
\vskip -0.45in
\epsfysize=20truecm \vbox{\epsfbox{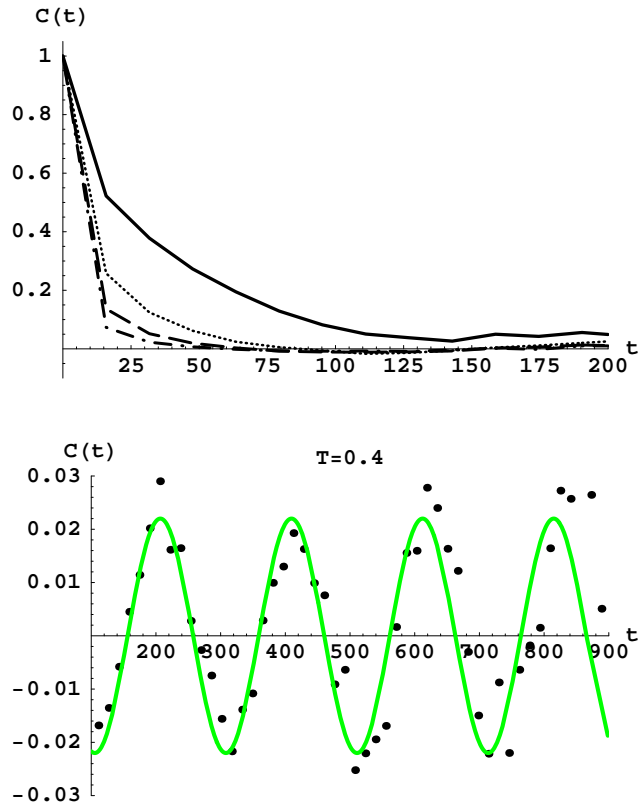}}
\end{center}
\vskip -2.5in
\caption{The upper plot shows the evolution of the temporal 
auto-correlation function (ACF) at early times and clearly indicates the exponential 
decay of the ACF. The four curves correspond to $T=0.2,0.4,0.6,0.8$ respectively from
top to bottom. The lower plot shows the oscillatory behaviour of the ACF at late
times for T=0.4. The solid line is a sinosoidal fit to the numerical data with 
an oscillating frequency $\omega=0.03$. Other parameter values are $\eta=0.03, 
\omega=0.03, A_0=0.1$.}
\label{Fig.7}
\end{figure}

\newpage

\vskip -0.35in
\begin{figure}[h]
\begin{center}
\leavevmode
\vskip -0.65in
\epsfysize=20truecm \vbox{\epsfbox{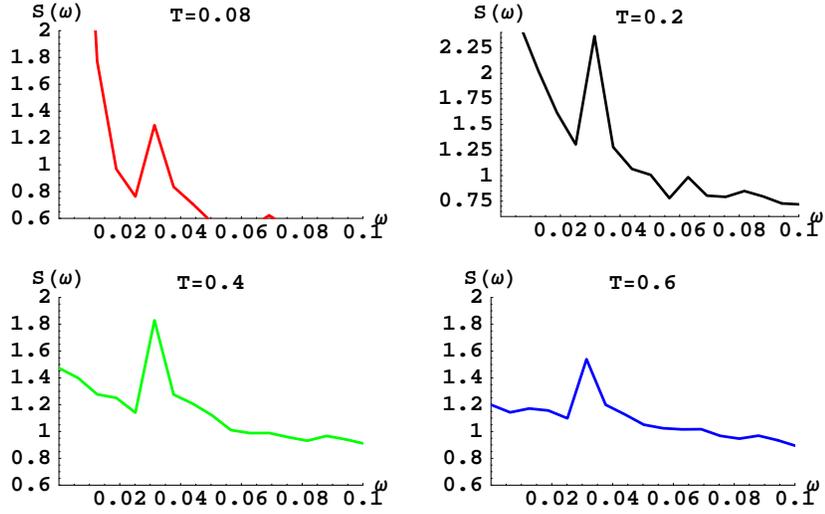}}
\end{center}
\vskip -3.0in
\caption{Plots of the power spectrum showing peaks at the resonant 
frequency $\omega_{sr}=0.03$. The largest peak is seen for $T=T_{sr}=0.2$ which 
corresponds to the temperature at which the resonance condition is satisfied for 
$\omega=0.03,\eta=0.03,A_0=0.1$.}
\label{Fig.8}
\end{figure}

\end{document}